\documentclass[conference]{IEEEtran}
\IEEEoverridecommandlockouts

\usepackage{cite}
\usepackage{amsmath,amssymb,amsfonts}
\usepackage{algorithmic}
\usepackage{graphicx}
\usepackage{textcomp}
\usepackage{xcolor}
\usepackage{subcaption}
\usepackage{tikz}
\usepackage{amsmath,amssymb,braket}
\usepackage{orcidlink}
\usetikzlibrary{decorations.pathreplacing,calc}

\def\BibTeX{{\rm B\kern-.05em{\sc i\kern-.025em b}\kern-.08em
    T\kern-.1667em\lower.7ex\hbox{E}\kern-.125emX}}
\begin{document}

\title{Detrimental Agnostic Entanglement: The Case Against Hardware-Efficient Ansätze for Combinatorial Optimization\\

}

\author{
    \IEEEauthorblockN{
        Tobias Rohe\IEEEauthorrefmark{1}\IEEEauthorrefmark{3}\orcidlink{0009-0003-3283-0586},
        Markus Baumann\IEEEauthorrefmark{1}\orcidlink{0009-0007-3575-1006},
        Federico Harjes Ruiloba\IEEEauthorrefmark{1}\IEEEauthorrefmark{2}\orcidlink{0009-0002-2283-0921}, 
        Philipp Altmann\IEEEauthorrefmark{1}\orcidlink{0000-0003-1134-176X}, \\
        Gerhard Stenzel\IEEEauthorrefmark{1}\orcidlink{0009-0009-0280-4911}, and 
        Claudia Linnhoff-Popien\IEEEauthorrefmark{1}\orcidlink{0000-0001-6284-9286}
    }
    \IEEEauthorblockA{
        \IEEEauthorrefmark{1}Institute for Computer Science, LMU Munich, 80538 Munich, Germany\\
    }
    \IEEEauthorblockA{
        \IEEEauthorrefmark{2}relAI – Konrad Zuse School of Excellence in Reliable AI, LMU Munich, 80799 Munich, Germany
    }
    \IEEEauthorblockA{
        \IEEEauthorrefmark{3}Email: tobias.rohe@ifi.lmu.de
    }
}

\maketitle

\begin{abstract}
Variational quantum algorithms (VQAs) for combinatorial optimization routinely employ entangling gates as a default design choice, yet the role of entanglement, in its amount and structure, remains poorly understood. This gap is particularly consequential for problems governed by diagonal Hamiltonians, whose ground states are classical product states and therefore require no entanglement in principle, raising the fundamental question of whether and how entangling gates help or hinder the variational search. We investigate this question for MaxCut by introducing two complementary control mechanisms that provide smooth, monotonic control over hardware-efficient ansatz (HEA) entanglement as quantified by the Meyer–Wallach measure $Q$, and by benchmarking against QAOA as a problem-structured reference. Tracking the entanglement trajectory $Q(t)$ throughout VQA training reveals that when the ansatz grants the optimizer indirect control over entanglement through its parameters, it consistently drives entanglement down. In line with this tendency, a fully separable ansatz outperforms all entangled hardware-efficient configurations, establishing a monotonic relationship: less problem-agnostic entanglement yields better performance. In contrast, QAOA, whose entanglement is structurally derived from the problem Hamiltonian, maintains high entanglement yet achieves competitive solution quality, demonstrating that entanglement structure, not merely quantity, determines its utility. These findings suggest that HEAs for diagonal Hamiltonians are inappropriate and that variational approaches to combinatorial optimization should prioritize problem-structured circuit designs.  
\end{abstract}


\section{Introduction}
Variational quantum algorithms (VQAs) have emerged as a leading paradigm for near-term quantum computation, with combinatorial optimization among their most prominent application domains~\cite{peruzzo2014variational, farhi2014quantum}. In this setting, a parameterized quantum circuit prepares trial states whose energy with respect to a cost Hamiltonian is minimized by a classical optimizer in a hybrid feedback loop. A central design choice in any such algorithm is the circuit ansatz, and in particular the role played by its entangling gates. Hardware-efficient ans\"atze (HEAs)~\cite{kandala2017hardware}, which compose layers of native single-qubit rotations and nearest-neighbor entangling gates, have become a widely adopted default due to their compatibility with current device topologies~\cite{nannicini2019performance, barkoutsos2020improving, amaro2022filtering, liu2022layer, miki2022variational, turati2023benchmarking, nakhl2024calibrating}. In these circuits, entangling gates, typically CNOT or CRZ ladders, are included because the hardware supports them, with little consideration of whether the entanglement they generate is beneficial for the problem at hand.

This practice rests on an implicit assumption: that entanglement, as the defining resource distinguishing quantum from classical computation, should generally aid the optimization process. Yet for an important class of combinatorial problems, those governed by diagonal Hamiltonians, including MaxCut, Max-SAT, and a broad range of QUBO formulations, the ground states are classical product states in the computational basis. Any entanglement present in the final variational state is, by definition, a deviation from the target. This raises a fundamental question that remains insufficiently addressed in the literature: does entanglement help or hinder the variational search for combinatorial optimization, and does the answer depend on the \emph{structure} of entanglement introduced?

Recent work has begun to probe this question from several angles. Studies have shown that product-state ans\"atze can be competitive with entangled circuits for certain combinatorial problems~\cite{diez2021quantum, rohe2024questionable}, that optimization performance can peak at intermediate entanglement levels for Hamiltonians with area-law entangled ground states~\cite{kim2022quantum}, and that entanglement requirements in QAOA are problem-dependent~\cite{nakhl2024calibrating}. However, a systematic investigation that (i)~provides direct experimental control over the amount of entanglement in the ansatz, (ii)~tracks how entanglement evolves dynamically during training, and (iii)~distinguishes between problem-agnostic and problem-structured entanglement is still lacking.

In this work, we introduce two complementary mechanisms for controlling entanglement in HEAs, progressive CNOT gate deletion and controlled-rotation parameter restriction, and validate that they provide smooth, monotonic modulation of circuit entanglement as measured by the Meyer--Wallach measure $Q$. We then conduct a systematic study on randomly generated MaxCut instances, tracking the entanglement trajectory $Q(t)$ throughout VQA training across multiple HEA configurations. To contextualize the HEA findings, we benchmark against QAOA, whose entangling structure is derived directly from the problem Hamiltonian. Our investigation is guided by two research questions:

\begin{itemize}
    \item[\textbf{RQ1:}] Can the amount of entanglement in HEAs be effectively and monotonically controlled?
    \item[\textbf{RQ2:}] How does entanglement evolve during variational training for MaxCut, and do its amount and structure determine solution quality?
\end{itemize}

\section{Background}
\subsection{MaxCut Problem}
Given an undirected graph $G=(V,E)$ with $n=|V|$ vertices, the Maximum Cut (MaxCut) problem asks for a bipartition of~$V$ into complementary sets $S$ and $\bar{S}$ that maximises the number of edges between vertices of both sets. Encoding the partition as a binary string $\mathbf{x}\in\{0,1\}^{n}$, where $x_v=0$ if $v\in S$ and $x_v=1$ otherwise, the cost function reads 
\begin{equation}
    C(\mathbf{x}) \;=\; \sum_{(u,v)\in E} x_u + x_v - 2\,x_u\,x_v\,.
    \label{eq:maxcut_cost}
\end{equation}

Under the substitution $x_v\to\tfrac{1}{2}(1-\sigma_v^z)$, this maps to a diagonal Ising Hamiltonian
\begin{equation}
    H_C \;=\; -\frac{|E|}{2}\,\mathbb{I}
             \;+\;\frac{1}{2}\sum_{(u,v)\in E}\sigma_u^z\,\sigma_v^z\,,
    \label{eq:maxcut_hamiltonian}
\end{equation}
whose ground-state eigenvector encodes the optimal cut and whose ground-state eigenvalue equals $-C_{\mathrm{opt}}$, so that the cut value is recovered as $C = -\langle H_C\rangle$. Eq.~\eqref{eq:maxcut_hamiltonian} differs from the conventional Ising form $\sum_{(u,v)\in E}\sigma_u^z\sigma_v^z$ only by a constant offset and an overall factor of $\tfrac{1}{2}$; in particular, the two share the same ground state. We retain the full form because our approximation ratios are computed from $\langle H_C\rangle$ via the identity $C=-\langle H_C\rangle$, which requires the offset and prefactor. Because $H_C$ is diagonal in the computational basis, the target state is a classical product state---a property that makes MaxCut a particularly interesting case for studying entanglement requirements. Any entanglement generated during the optimisation must ultimately vanish in the solution, yet may still be beneficial as a transient resource during the search.

MaxCut is NP-hard~\cite{Karp1972Reducibility}; the best known classical guarantee is the Goemans--Williamson semidefinite-programming bound with an approximation ratio of $\alpha_{\mathrm{GW}}\approx 0.8786$~\cite{goemans1995improved}, used here as an orientation for solution quality.

\subsection{Variational Quantum Eigensolver}
The Variational Quantum Eigensolver (VQE), introduced by Peruzzo et~al.~\cite{peruzzo2014variational}, is a hybrid quantum--classical algorithm that approximates the ground-state energy of a Hamiltonian $H_C$ by minimising the expectation value
\begin{equation}
    E(\boldsymbol{\theta})
      \;=\; \langle 0|^{\otimes n}\,
            U^{\dagger}(\boldsymbol{\theta})\,H_C\,
            U(\boldsymbol{\theta})\,|0\rangle^{\otimes n}
\label{eq:vqe_cost}
\end{equation}
over parameters $\boldsymbol{\theta}$ of a parameterised quantum circuit (ansatz) $U(\boldsymbol{\theta})$. The quantum processor prepares the trial state $|\psi(\boldsymbol{\theta})\rangle = U(\boldsymbol{\theta})|0\rangle^{\otimes n}$ and estimates $E(\boldsymbol{\theta})$ via repeated measurement, while a classical optimiser updates $\boldsymbol{\theta}$ in a closed loop.

In contrast to the Quantum Approximate Optimisation Algorithm~\cite{farhi2014quantum}, whose phase-separation layers are derived from the problem Hamiltonian $H_C$ and thus encode the problem structure directly into the circuit, VQE is typically paired with a HEA that composes layers of native single-qubit rotations and nearest-neighbour entangling gates in a problem-agnostic fashion~\cite{kandala2017hardware}. HEAs offer flexibility and low gate overhead, yet their problem-agnostic structure means that the amount of entanglement they generate is not adapted to the target state.

\subsection{Quantum Approximate Optimisation Algorithm}
The Quantum Approximate Optimisation Algorithm (QAOA), introduced by Farhi et al.~\cite{farhi2014quantum}, constructs a variational ansatz directly from the problem Hamiltonian. Starting from the uniform superposition $|+\rangle^{\otimes n}$, the circuit alternates $p$ layers of two unitaries: a phase-separation operator derived from the cost Hamiltonian $H_C$ and a mixing operator based on a transverse-field Hamiltonian $H_M = \sum_{i=1}^{n} \sigma_i^x$. The trial state after $p$ layers reads
\begin{equation}
    |\boldsymbol{\gamma}, \boldsymbol{\beta}\rangle
      \;=\; \prod_{l=1}^{p}\,
            e^{-i\beta_l H_M}\,
            e^{-i\gamma_l H_C}\,
            |+\rangle^{\otimes n}\,,
    \label{eq:qaoa_state}
\end{equation}
where $\boldsymbol{\gamma} = (\gamma_1, \dots, \gamma_p)$ and $\boldsymbol{\beta} = (\beta_1, \dots, \beta_p)$ are optimised classically to minimise $\langle \boldsymbol{\gamma}, \boldsymbol{\beta} | H_C | \boldsymbol{\gamma}, \boldsymbol{\beta} \rangle$. For MaxCut, the phase-separation unitary decomposes into two-qubit $ZZ$-rotations acting on each edge,
\begin{equation}
    e^{-i\gamma H_C}
      \;=\; e^{i\gamma\frac{|E|}{2}}\,\prod_{(u,v)\in E}\,
            e^{-i\frac{\gamma}{2}\,\sigma_u^z\sigma_v^z}\,,
    \label{eq:qaoa_phase_sep}
\end{equation}
so that the entangling topology of each QAOA layer is determined entirely by the graph $G$. This stands in contrast to HEAs, where the entangling topology is fixed by the device layout. The number of variational parameters in QAOA is $2p$, independent of the problem size $n$, which makes the circuit compact but limits the optimiser's flexibility. In the limit $p\to\infty$, QAOA recovers the adiabatic algorithm and is guaranteed to find the ground state~\cite{farhi2014quantum}; at finite $p$, the achievable approximation quality depends on both the circuit depth and the problem structure.

\section{Related Work}
Gross et al. demonstrated that quantum states can be too entangled to serve as useful computational resources~\cite{gross2009most}. They established that the fraction of useful $n$-qubit pure states is bounded by $\exp(-n^2)$, rendering computational universality an exceedingly rare property in Hilbert space. This work established a foundational principle: entanglement must be present in the right dose to yield computational utility. More is therefore not always better.

Subsequent work has explored how this principle manifests in variational quantum optimization. Díez-Valle et al.~\cite{diez2021quantum} investigated the relationship between entanglement structure and problem topology. By constructing ansätze with entangling layers that mirror the connectivity of the underlying QUBO graph, their circuits implicitly scale entanglement with problem density where sparse graphs induce fewer entangling gates, while dense graphs require more. Their results demonstrate that such topology-adapted entanglement yields advantages for low-density problems, but this benefit vanishes as graph density increases. Notably, product-state ansätze, circuits that entirely avoid entanglement, achieved competitive performance, particularly when combined with conditional value-at-risk cost functions.

Kim and Oz~\cite{kim2022quantum} provided a geometric perspective on why intermediate entanglement may be optimal. By systematically reducing entangling capability through stochastic dropout of two-qubit gates, they demonstrated that optimization performance peaks within an intermediate entanglement regime. Their Hessian eigenspectrum analysis revealed that low-entangling circuits develop favorable curvature properties resembling over-parameterized classical networks, enabling faster convergence. Although their study focused on the transverse-field Ising model with area-law entangled ground states, the finding that intermediate entanglement optimizes trainability resonates with combinatorial optimization, where ground states are classical product states. Nakhl et al.~\cite{nakhl2024calibrating} corroborated these findings using tensor network simulations, showing that QAOA entanglement requirements for MaxCut are problem-dependent: sparse and structured graphs require less entanglement than dense graphs, which is in line with Díez-Valle et al.~\cite{diez2021quantum}.

Wang et al.~\cite{wang2024entanglement} addressed the entanglement control problem from a circuit design perspective, proposing an entanglement-variational HEA in which two-qubit entanglers are parameterized rather than fixed. While entanglement can be rapidly generated within a few circuit layers, excess entanglement cannot be efficiently removed during optimization. By making entanglers variational, their ansatz can learn to adjust entanglement to the appropriate level. Although their work targets quantum many-body and chemistry problems, the underlying principle that entanglement must match problem structure, directly motivates our investigation.

Recently, the limitations of HEAs have also been formalized theoretically. Leone et al.~\cite{leone2024practical} rigorously demonstrated that the trainability and classical simulability of HEAs are strictly governed by the entanglement of the input state. Crucially, they proved that for VQAs starting from product states, the standard initialization for combinatorial optimization, shallow HEAs remain efficiently classically simulable, thereby ruling out their use for achieving quantum advantage in these settings. Our experimental tracking of entanglement dynamics directly corroborates these theoretical limitations. Our work differs in that we measure and manipulate entanglement directly, rather than only through variational entanglers, targeting combinatorial problems whose classical ground states are product states.

\section{Methodology}
Our experimental design treats entanglement as a controlled variable rather than a fixed circuit property. We introduce two complementary mechanisms that modulate entanglement in HEAs — one structural, one parametric — and quantify their effect via the Meyer–Wallach measure $Q$. These tools are then deployed in a systematic VQE study on randomly generated MaxCut instances, with QAOA serving as a problem-structured reference. The following subsections detail each component in turn.

\subsection{Entanglement Control Mechanisms}
\label{sec:entanglement_control}

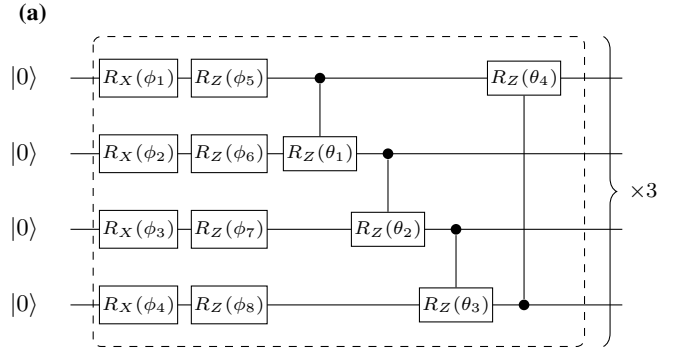
\begin{figure}[t]
\centering
\begin{tikzpicture}[
    gate/.style={draw, fill=white, minimum width=0.9cm, minimum height=0.5cm, 
                 font=\scriptsize, inner sep=1pt},
    rzbox/.style={draw, fill=white, minimum width=0.85cm, minimum height=0.45cm, 
                  font=\scriptsize, inner sep=1pt},
    ctrl/.style={fill=black, circle, inner sep=0pt, minimum size=4pt},
    targ/.style={draw, circle, inner sep=0pt, minimum size=8pt, line width=0.4pt,
        path picture={
            \draw[line width=0.4pt] 
                (path picture bounding box.south) -- (path picture bounding box.north);
            \draw[line width=0.4pt] 
                (path picture bounding box.west) -- (path picture bounding box.east);
        }
    }
]

\def\yoff{-5.0}

\node[font=\small\bfseries] at (0.3, 0.85+\yoff) {(a)};

\foreach \i in {0,...,3} {
    \pgfmathsetmacro{\ypos}{\yoff - \i}
    \node[left, font=\small] at (0.5, \ypos) {$\ket{0}$};
    \draw (0.8, \ypos) -- (8.1, \ypos);
}

\node[gate] at (1.7, \yoff)   {$R_X(\phi_1)$};
\node[gate] at (1.7, \yoff-1) {$R_X(\phi_2)$};
\node[gate] at (1.7, \yoff-2) {$R_X(\phi_3)$};
\node[gate] at (1.7, \yoff-3) {$R_X(\phi_4)$};

\node[gate] at (2.9, \yoff)   {$R_Z(\phi_5)$};
\node[gate] at (2.9, \yoff-1) {$R_Z(\phi_6)$};
\node[gate] at (2.9, \yoff-2) {$R_Z(\phi_7)$};
\node[gate] at (2.9, \yoff-3) {$R_Z(\phi_8)$};

\node[ctrl] (c2a) at (4.1, \yoff) {};
\node[rzbox] (t2a) at (4.1, \yoff-1) {$R_Z(\theta_1)$};
\draw (c2a) -- (t2a);

\node[ctrl] (c2b) at (5.0, \yoff-1) {};
\node[rzbox] (t2b) at (5.0, \yoff-2) {$R_Z(\theta_2)$};
\draw (c2b) -- (t2b);

\node[ctrl] (c2c) at (5.9, \yoff-2) {};
\node[rzbox] (t2c) at (5.9, \yoff-3) {$R_Z(\theta_3)$};
\draw (c2c) -- (t2c);

\node[ctrl] (c2d) at (6.8, \yoff-3) {};
\node[rzbox] (t2d) at (6.8, \yoff) {$R_Z(\theta_4)$};
\draw (c2d) -- (t2d);

\draw[dashed, rounded corners=3pt] (1.1, 0.55+\yoff) rectangle (7.6, -3.55+\yoff);

\draw[decorate, decoration={brace, amplitude=5pt}] 
    (7.85, 0.55+\yoff) -- (7.85, -3.55+\yoff) 
    node[midway, right=6pt, font=\footnotesize] {$\times 3$};
\end{tikzpicture}
\caption{HEA architectures. The circuit illustrates a single layer (repeated $L=3$ times) consisting of parameterized single-qubit rotations $R_X(\phi), R_Z(\phi)$ followed by an entangling block. \textbf{(a)} In the \textit{crz\_ring} configuration (shown), entanglement is mediated by controlled-$R_Z$ gates with variational parameters $\theta$, allowing the optimizer to tune entangling strength. \textbf{(b)} The \textit{cnot\_ring} variant (not explicitly pictured) replaces these with fixed CNOT gates, removing $\theta$ from the parameter set. This separation of $\phi$ and $\theta$ enables independent initialization strategies and the controlled study of entangling resources (see Sec.~\ref{sec:experimental_protocol}).}
\label{fig:circuits}
\end{figure}

We employ two HEAs that share the same single-qubit structure with three layers of $R_X$ and $R_Z$ rotation ladders, but differ in their entangling blocks: circuit \textit{cnot\_ring} uses a ring of CNOT gates, while circuit \textit{crz\_ring} uses a ring of controlled-$R_Z$ (CRZ) gates. While our study utilizes $12$-qubits, Fig.~\ref{fig:circuits} illustrates the \textit{crz\_ring} structure using an exemplary four-qubit instance.

Each circuit, \textit{cnot\_ring} and \textit{crz\_ring}, is paired with a dedicated entanglement control mechanism that acts exclusively on the entangling block, leaving all single-qubit gates and their parameters unaffected. This provides one \emph{structural} (for the \textit{cnot\_ring}) and one \emph{parametric} (for the \textit{crz\_ring}) knob for modulating entanglement.

\textit{CNOT-Deletion.} Inspired by the stochastic gate-dropout of Kim and Oz~\cite{kim2022quantum}, we introduce a deletion factor $\delta \in [0.0,1.0]$ that specifies the fraction of CNOT gates to be removed from circuit \textit{cnot\_ring}. The number of deleted gates is $k = \lfloor \delta \cdot N_{\mathrm{CNOT}} \rfloor$, selected uniformly at random across all layers prior to optimization and held fixed throughout training. The limits $\delta=0$ and $\delta=1$ correspond to the full entangling circuit and a product-state ansatz, respectively.

\textit{Parameter-Restriction.} Since a CRZ$(\theta)$ gate produces no entanglement at $\theta=0$ and maximal entanglement at $|\theta|=\pi$, restricting the accessible range of $\theta$ should provide continuous control over the circuit's \textit{crz\_ring} entangling strength. A restriction factor $\rho \in [0.0,1.0]$ constrains each CRZ parameter to $\theta \in [-(1-\rho)\pi,\, +(1-\rho)\pi]$, so that $\rho=0$ imposes no restriction and $\rho=1$ yields a product-state ansatz.

To enforce this parameter restriction, we implement two strategies. \textit{Bound-constrained optimization} supplies the COBYLA optimizer with explicit inequality constraints on the CRZ parameter subset, restricting every entangling parameter to $\theta_j \in [-(1-\rho)\pi,\, +(1-\rho)\pi]$ throughout the entire optimization trajectory. At the limiting case $\rho = 1$, the permitted range collapses to $\theta_j = 0$, rendering every CRZ gate an identity operation. To avoid retaining dead parameters that would present the optimizer with uninformative flat directions, the CRZ gates are removed from the circuit entirely in this case and the corresponding entries are dropped from the optimization variable vector, so that the $\rho = 1$ setting represents a product-state ansatz.
The second strategy is a smooth \textit{tanh remapping}, where each CRZ gate receives an unconstrained parameter $\xi \in \mathbb{R}$ mapped to the effective angle via
\begin{equation}
    \theta(\xi) = (1 - \rho)\,\pi \cdot \tanh\!\left(\frac{\xi}{(1 - \rho)\,\pi}\right),
    \label{eq:tanh_remap}
\end{equation}
which guarantees $\theta \in [-(1-\rho)\pi,\, +(1-\rho)\pi]$ while ensuring that the gradient $\mathrm{d}\theta/\mathrm{d}\xi$ at the origin remains unity for all~$\rho$, preventing the optimizer's sensitivity from being artificially suppressed at high restriction levels. 
In our experiments, both enforcement strategies yield comparable optimization performance; we report results obtained with bound-constrained enforcement throughout, as it is the natural formulation for COBYLA, which natively supports inequality constraints.

\subsection{The Meyer-Wallach Measure}
\label{sec:meyer_wallach}

To quantify the entanglement present in our circuit states, we employ the Meyer--Wallach measure $Q$, a global multipartite entanglement measure introduced for systems of $n$ qubits~\cite{meyer2002global}. The measure can be expressed as the average single-qubit linear entropy: for an $n$-qubit pure state $|\psi\rangle$ with single-qubit reduced density matrices $\rho_i = \mathrm{tr}_{\bar{i}}(|\psi\rangle\langle\psi|)$, it is defined as
\begin{equation}
    Q(|\psi\rangle) = \frac{2}{n}\sum_{i=1}^{n}\bigl(1 - \mathrm{tr}(\rho_i^2)\bigr).
    \label{eq:meyer_wallach}
\end{equation}
Each term $1 - \mathrm{tr}(\rho_i^2)$ quantifies the mixedness of qubit $i$ when the remaining qubits are traced out; a qubit that is entangled with the rest of the system appears mixed in isolation, whereas an unentangled qubit remains pure. The measure is bounded between $Q = 0$ for product states and $Q = 1$ when every single-qubit reduced state is maximally mixed. By averaging over all qubits, $Q$ provides a single scalar that captures the overall degree of entanglement in the system, making it well-suited for tracking how our entanglement control mechanisms modulate the circuit state throughout the optimization process, however, it makes no statements about the structure of entanglement.
A detailed discussion of this formulation can be found in~\cite{brennen2003observable}.

\subsection{Problem Instances}
We evaluate all ansätze on MaxCut instances defined by Erd\H{o}s--R\'enyi random graphs $G(n,p)$ with $n=12$ vertices (one qubit per vertex). The primary benchmark uses $p=0.5$, with twenty graphs generated from deterministic seeds. For the edge-density analysis, independent sets of twenty graphs are generated at $p \in \{0.2, ... , 0.8\}$. For each graph, five independent optimization runs are performed with distinct parameter initialization seeds, producing $100$ (graph, seed) pairs per ansatz configuration. The exact maximum cut for every instance is obtained by brute-force enumeration of all $2^{12}$ bitstrings. All quantum simulations are performed via noiseless statevector simulation.

\subsection{Experimental Protocol}
\label{sec:experimental_protocol}
The separation of circuit parameters into single-qubit rotations $\boldsymbol{\phi}$ and entangling gate parameters $\boldsymbol{\theta}$ (cf.\ Fig.~\ref{fig:circuits}) enables independent initialization strategies for each group. We consider two initialization regimes: \textit{random}, where parameters are sampled uniformly from $[-\pi, +\pi]$, and \textit{near\_zero}, where parameters are sampled uniformly from $[-0.1, +0.1]$. For the \textit{cnot\_ring} circuit, which has no entangling parameters, only $\boldsymbol{\phi}$ is initialized (two configurations). For the \textit{crz\_ring} circuit, $\boldsymbol{\phi}$ and $\boldsymbol{\theta}$ are initialized independently, yielding four combinations. Setting $\delta = 1$ in the \textit{cnot\_ring} circuit removes all entangling gates and serves as a \textit{product-state} baseline. In addition, QAOA is evaluated at depths $p \in \{3, 5, 7\}$ using a warm-ramp initialization in which $\gamma_l$ increases and $\beta_l$ decreases linearly with layer index, with small Gaussian noise ($\sigma = 0.05$).

Unless otherwise specified, all circuits are optimized by minimizing $\langle \psi(\boldsymbol{\theta}) | H_C | \psi(\boldsymbol{\theta}) \rangle$ with the gradient-free COBYLA optimizer (initial trust-region radius $0.5$) for $500$ iterations. All quantum simulations use noiseless statevector simulation. We quantify solution quality by the approximation ratio $\alpha = C(\mathbf{x}^*) / C_{\mathrm{opt}}$, where $C_{\mathrm{opt}}$ is the exact maximum cut.

\section{Results}
\subsection{Validation of Entanglement Control (RQ1)}

\begin{figure*}[t]
    \centering
    \includegraphics[width=0.48\textwidth]{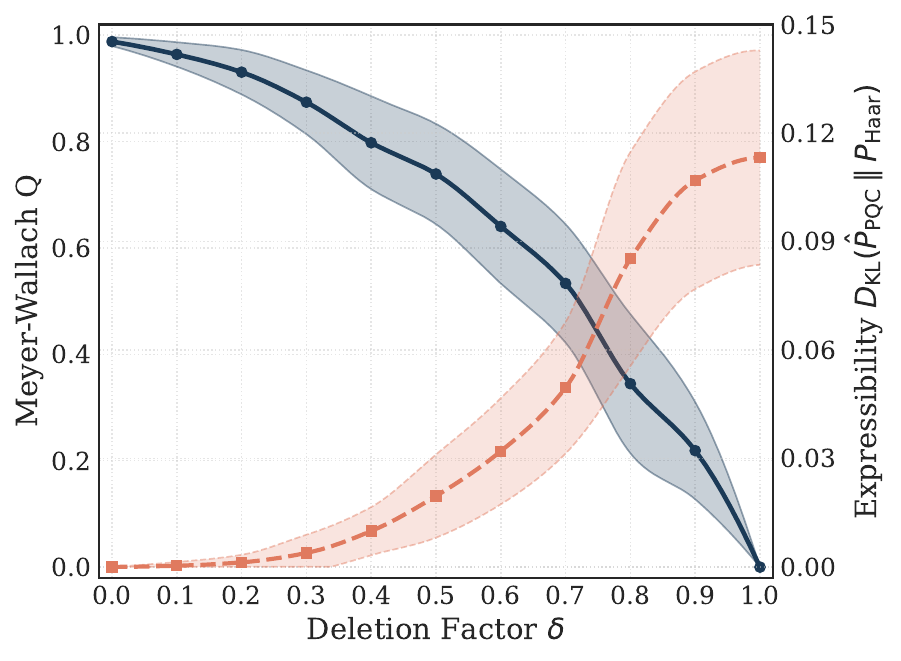}%
    \hfill
    \includegraphics[width=0.48\textwidth]{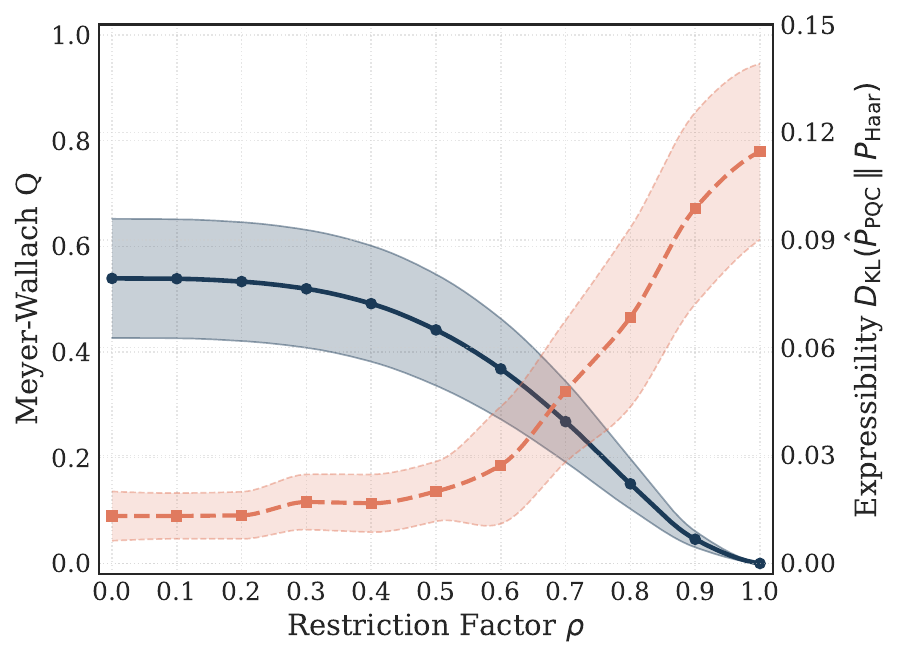}\\[3pt]
    \includegraphics[width=0.55\textwidth]{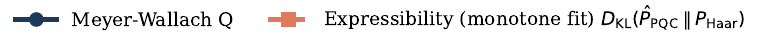}
    \caption{Validation of the two entanglement control mechanisms on the HEAs (averaged over random parameter initializations; shaded regions indicate one standard deviation). Left: Progressive CNOT gate deletion (where $\delta = 0$ retains all CNOTs and $\delta = 1$ removes all). Right: Controlled-rotation (CRot) parameter restriction (where $\rho = 0$ applies no restriction and $\rho = 1$ restricts all CRot parameters to zero). 
    }
    \label{fig:do_re_factors}
\end{figure*}

Before investigating the relationship between entanglement and optimization performance, we must verify that our two mechanisms to control for the amount of entanglement provide reliable and monotonic control. We therefore evaluate $Q$ and the circuit expressibility $\mathrm{Expr}$ as functions of the deletion factor $\delta \in [0,1]$ and the restriction factor $\rho \in [0,1]$, respectively, over ensembles of randomly sampled parameter vectors $\boldsymbol{\phi}$ and $\boldsymbol{\theta}$. Results are shown in Fig.~\ref{fig:do_re_factors}.

For both mechanisms, $Q$ decreases monotonically from near-maximal entanglement to the product-state limit $Q = 0$ as the control parameter increases, confirming that each intervention spans the full entanglement range with no discontinuities or inversions. Under CNOT deletion (Fig.~\ref{fig:do_re_factors}, left), $Q$ decreases smoothly from $Q \approx 0.98$ at $\delta = 0$ to $Q = 0$ at $\delta = 1$, exhibiting the steepest descent in the range $\delta \in [0.8, 1.0]$. Concurrently, the expressibility decreases (i.e., the KL divergence from the Haar-random fidelity distribution grows), indicating that entangling gates are the primary source of expressibility in HEAs. The confidence bands remain narrow throughout, reflecting consistent behavior across parameter initializations. As the approach is similar to Kim and Oz~\cite{kim2022quantum} we also observe an identical behavior.

Under CRot restriction (Fig.~\ref{fig:do_re_factors}, right), $Q$ starts at a lower baseline of $Q \approx 0.50$ — expected because controlled-rotation gates with unrestricted but finite-range parameters generate less average entanglement than fixed CNOT gates — and then exhibits a pronounced plateau for $\rho \in [0, 0.4]$, during which $Q$ remains effectively constant. Only beyond $\rho \approx 0.5$ does entanglement decline sharply, reaching $Q = 0$ at $\rho = 1$. This plateau has a clear operational interpretation: because the CRot gate interpolates continuously between the identity ($\theta = 0$) and a maximally entangling operation ($\theta = \pm\pi$), compressing the rotation angles by up to around 50\% (effectively $\theta = \pm 1/2 \pi$) still leaves sufficient parametric freedom to produce high levels of entanglement. The plateau thus delineates a saturation regime in which the available entangling capacity exceeds what random parameter sampling can exploit, additional angular freedom yields no further entanglement. Only when the angular budget is restricted beyond this saturation threshold does the effective entangling power of the circuit begin to degrade.

Two conclusions follow from these results. First, both $\delta$ and $\rho$ constitute valid and complementary entanglement control knobs: deletion modifies the circuit structure (gate-level, discrete), whereas restriction modulates the circuit parametrization (topology-preserving for $\rho < 1$, continuous). Second, as entanglement decreases under both mechanisms, the KL divergence $D_{\mathrm{KL}}\bigl(P_{\text{circuit}} \| P_{\text{Haar}}\bigr)$ increases monotonically, indicating that the circuit's state distribution moves further from the Haar-random reference — i.e., the circuit becomes less expressible. Reducing entanglement therefore simultaneously restricts the volume of Hilbert space explored to the ansatz. Consequently, any performance gains observed at reduced entanglement levels cannot be attributed to increased expressibility.

\subsection{Entanglement Dynamics During Training (RQ2)}
\label{sec:results_rq2}
We now turn to the central question of this work: How does entanglement evolve during variational training for MaxCut, and does its amount determine solution quality? We address this by analyzing (i)~the entanglement trajectory $Q(t)$ throughout training, (ii)~the aggregate solution quality across all problem instances, and (iii)~the dependence of performance on graph structure, specifically edge density and the number of global optima.

\subsubsection{Entanglement Trajectories During Training}

\begin{figure*}[t]
    \centering
    \subfloat[Hardware-efficient ans\"atze.\label{fig:q_trajectories_hea}]{%
        \includegraphics[width=0.49\textwidth]{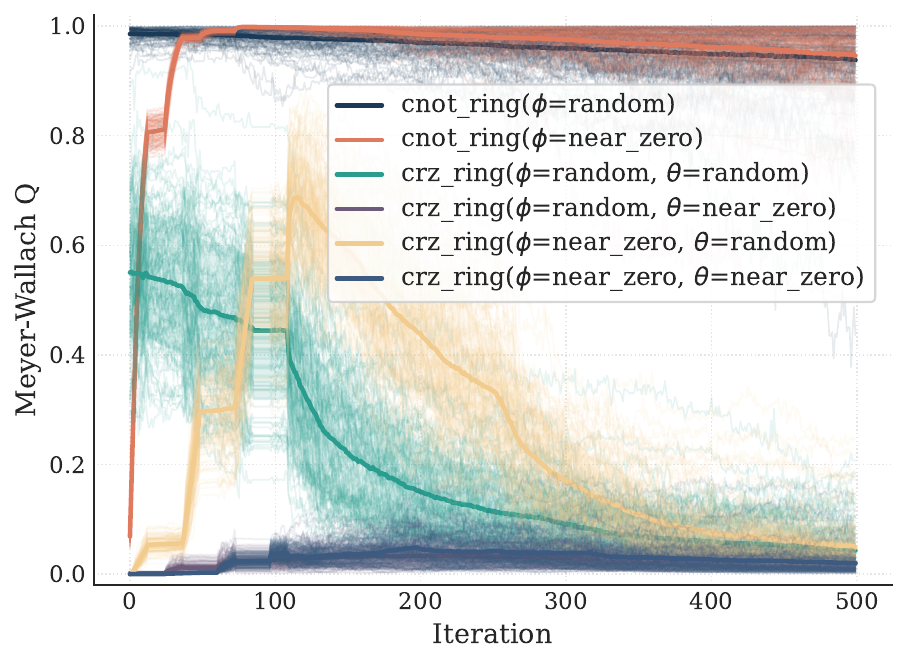}}
    \hfill
    \subfloat[QAOA ans\"atze ($p=3, 5, 7$).\label{fig:q_trajectories_qaoa}]{%
        \includegraphics[width=0.49\textwidth]{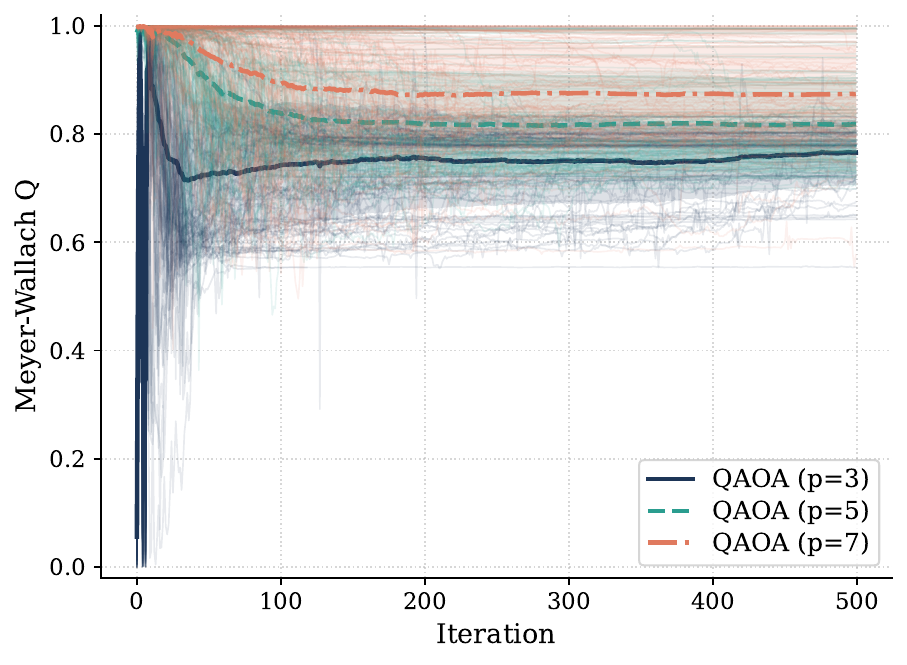}}
    \caption{Meyer--Wallach entanglement measure $Q$ as a function of optimizer iterations. Faint lines show individual runs; bold lines indicate the per-configuration mean. 
    }
    \label{fig:q_trajectories}
    
\end{figure*}

Figure~\ref{fig:q_trajectories} shows the Meyer--Wallach measure $Q$ over the course of VQE and QAOA optimization for all ansatz configurations. Each circuit is evaluated on $100$ distinct optimization trajectories ($5$ seeds $\times$ $20$ independently generated Erd\H{o}s--R\'enyi graphs); faint lines trace individual runs and bold lines their per-configuration mean.

\paragraph{CNOT-based Ans\"atze.}
Because the CNOT gates impose a \emph{static} entanglement topology, the entanglement dynamics of these circuits are governed by the single-qubit rotation parameters. Under \textit{cnot\_ring($\phi=\text{random}$)}, uniform initialization immediately drives the state toward near-maximal entanglement ($Q \approx 1.0$), which remains largely stagnant—settling around $0.95$ after $500$ iterations. This limited adaptability is a direct consequence of the rigid entanglement structure: the optimizer can redistribute correlations through the rotation gates, but it cannot \emph{remove} entanglement that the ladder of CNOT-gates inherently generates.

The \textit{cnot\_ring($\phi=near\_zero$)} variant starts from a near-product state ($Q \approx 0$), as intended by the near-zero initialization. During training, $Q$ rises fast as the optimizer tunes the rotation angles away from their initial zero-values, necessarily generating entanglement through the ever-present CNOT-gates. By iteration $500$, the entanglement has grown substantially ($Q \approx 0.95$), illustrating that the static entangling layer \emph{forces} entanglement production once the rotation parameters depart from the identity limit. The sharp increase observed here contrasts with the CRot-based circuits and highlights a fundamental limitation of CNOT-only ans\"atze: the optimizer has no mechanism to independently suppress entanglement.

\paragraph{CRot-based ans\"atze with near-zero entanglement initialization.}
The configurations \textit{crz\_ring($\phi=random, \theta=near\_zero$)} and \textit{crz\_ring($\phi=near\_zero, \theta=near\_zero$)} both start at $Q \approx 0$ and remain near zero throughout training, with only a marginal transient increase during the first $\sim 100$ iterations that is suppressed as the optimizer converges. The \textit{crz\_ring($\phi=random, \theta=near\_zero$)} case is particularly telling: the single-qubit rotations are already randomly initialized and span a diverse region of product-state space, so the optimizer has access to entangling degrees of freedom but no incentive to use them. The brief $Q$ fluctuation is consistent with COBYLA's initial simplex construction perturbing the CRot parameters away from zero, but the optimizer rapidly corrects this and solves the problem entirely within the product-state manifold. The close agreement between the two variants confirms that the single-qubit initialization has negligible influence on the entanglement dynamics when the entangling layer is itself initialized near the identity.

\paragraph{CRot-based ans\"atze with random entanglement initialization.}
The two configurations with randomly initialized CRot parameters exhibit distinct initial conditions: \textit{crz\_ring($\phi=random,\theta=random$)} starts at $Q \approx 0.55$, as the random single-qubit rotations immediately activate the entangling layer, whereas \textit{crz\_ring($\phi=near\_zero,\theta=random$)} starts at $Q \approx 0$ because the near-zero rotations keep the state close to $|0\rangle^{\otimes n}$, where the CRZ controls do not fire despite their random parameterization. The latter rises sharply to $Q \approx 0.8$ as the single-qubit parameters grow and activate the entangling layer before descending to $Q \approx 0.05$, while the former descends monotonically from the start. Despite these different transient profiles, both converge to $Q \approx 0.05$ by iteration $500$---approximately the same low-entanglement regime reached by the near-zero-initialized CRot variants. The optimizer thus drives the circuit toward the product-state manifold regardless of initial entanglement level, confirming that the energy landscape for these diagonal Hamiltonians consistently favors separable solutions.

\paragraph{QAOA.}
The QAOA variants (Figure~\ref{fig:q_trajectories_qaoa}) start at a high entanglement level ($Q$ close to $1.0$) at the very first iterations and reduce to a stable plateau near $Q \in [0.7, 0.9]$ within the first $\sim200$ iterations. A notable feature is the high variance and rapid oscillation of $Q$ observed for $p=3$ during the first $\sim20$ iterations, where the measure swings between $0.0$ and $1.0$. This behavior is likely an artifact of the COBYLA optimizers initial simplex construction; the large sampling steps required to build the local linear model, combined with the extreme sensitivity of the alternating unitary structure to its few variational parameters ($2p=6$), result in dramatic shifts in the state's entanglement. As the trust region contracts the trajectory stabilizes at a relatively high $Q$ plateau. This stands in marked contrast to the CRot-based circuits, where the entanglement can be flexibly driven to much lower levels. This observation of QAOA entanglement behavior is consistent with findings of Dupont et al.~\cite{dupont2022entanglement}.

\paragraph{Summary of dynamical regimes.}
The entanglement trajectories reveal a key structural distinction: CNOT-based ans\"atze impose \emph{rigid} entanglement that the optimizer can modulate only indirectly through the rotation parameters, whereas CRot-based ans\"atze provide more \emph{flexible} entanglement control, enabling the optimizer to actively increase or decrease $Q$ as can be seen in the $Q$-trajectory. Across all configurations with flexible entanglement control, the optimizer shows a consistent tendency to reduce entanglement during longer training. QAOA occupies its own position: its entanglement is structurally imposed and exhibits, like the \textit{cnot\_ring} circuits, a high level of $Q$, however, its entanglement is derived from the problem-instance at hand, not in a hardware-efficient way.

\subsubsection{Aggregated Solution Quality}
\begin{figure*}[t]
    \centering
    \subfloat[Approximation ratios by configuration.\label{fig:mean_approx_ratio}]{%
        \includegraphics[width=0.48\textwidth]{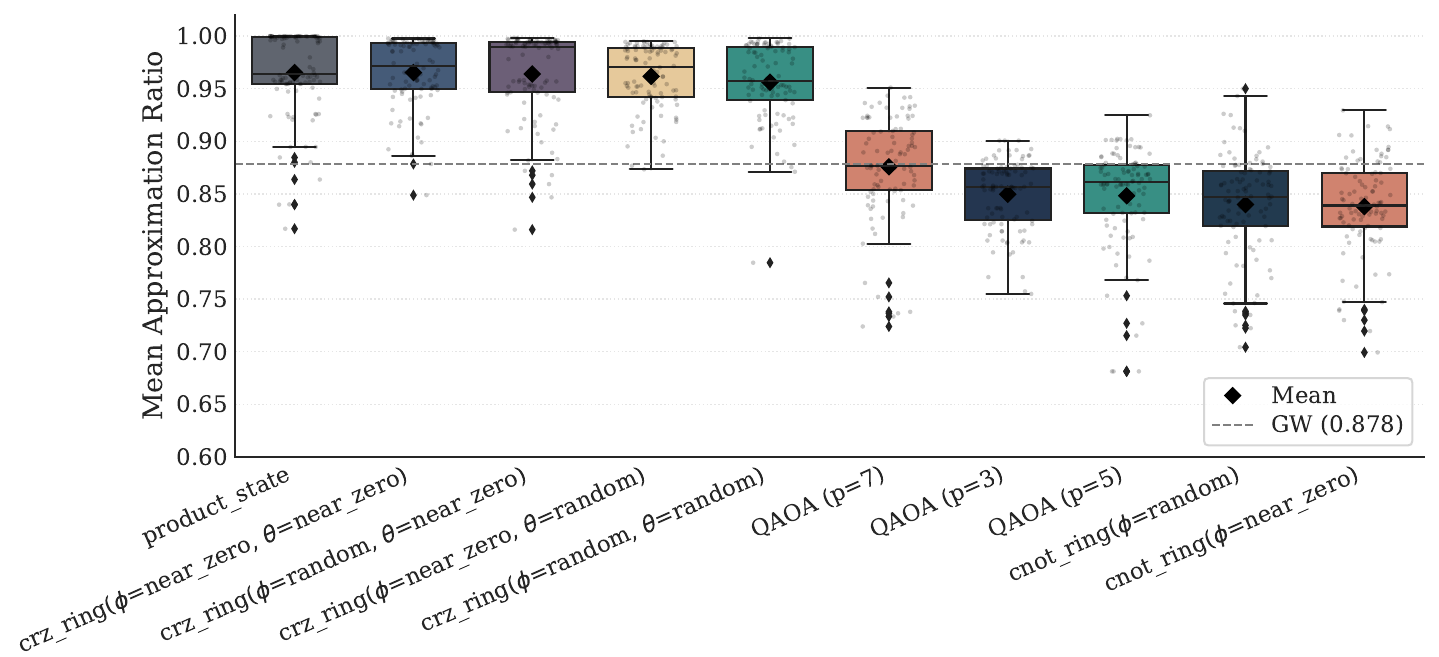}}
    \hfill
    \subfloat[QAOA mean vs.\ most probable bitstring quality.\label{fig:qaoa_bitstring}]{%
        \includegraphics[width=0.48\textwidth]{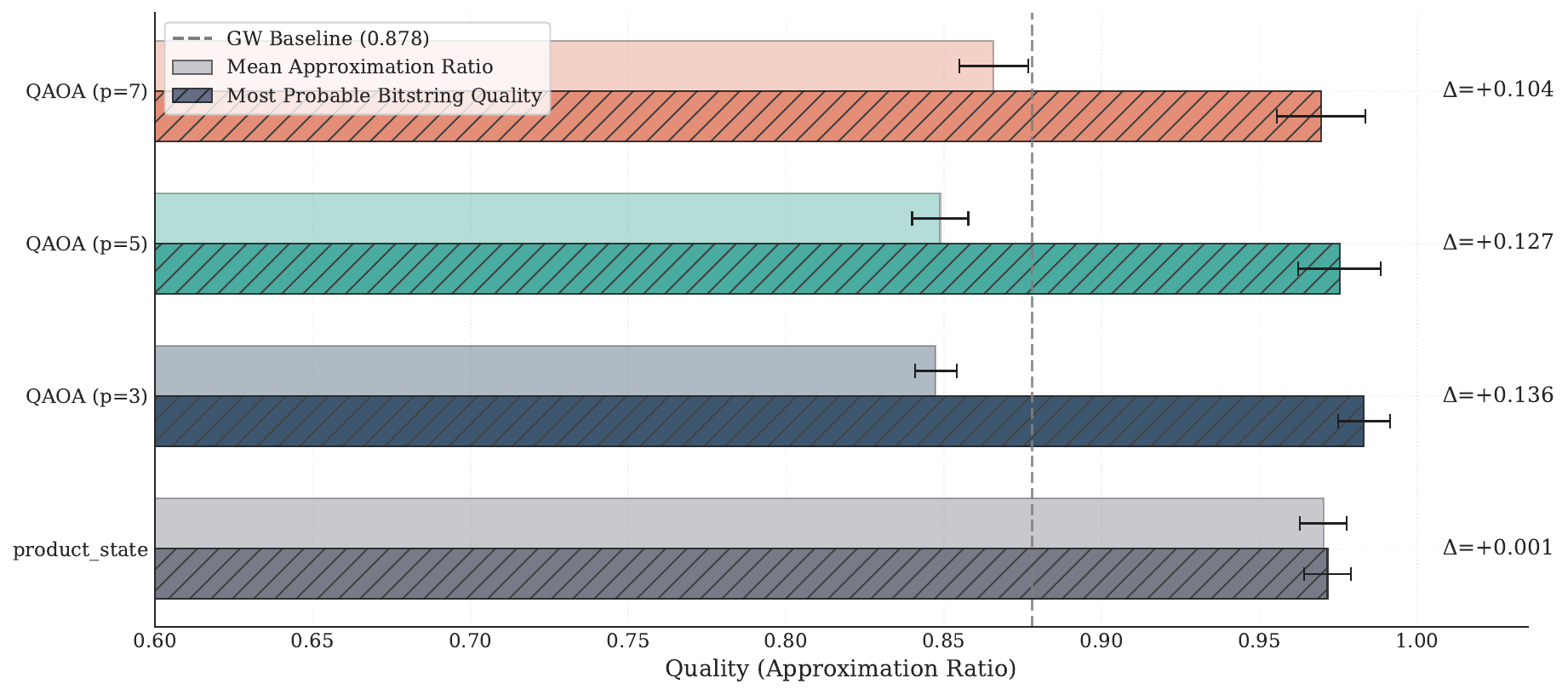}}\\[6pt]
    \subfloat[Final $Q$ vs.\ quality.\label{fig:corr_final}]{%
        \includegraphics[width=0.32\textwidth]{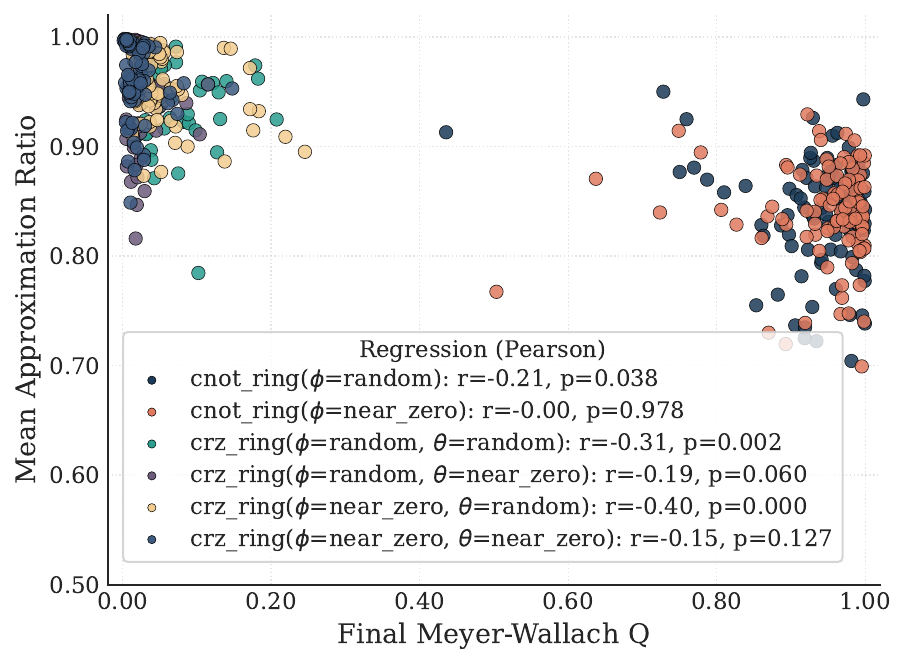}}
    \hfill
    \subfloat[Peak $\max_t Q(t)$ vs.\ quality.\label{fig:corr_max}]{%
        \includegraphics[width=0.32\textwidth]{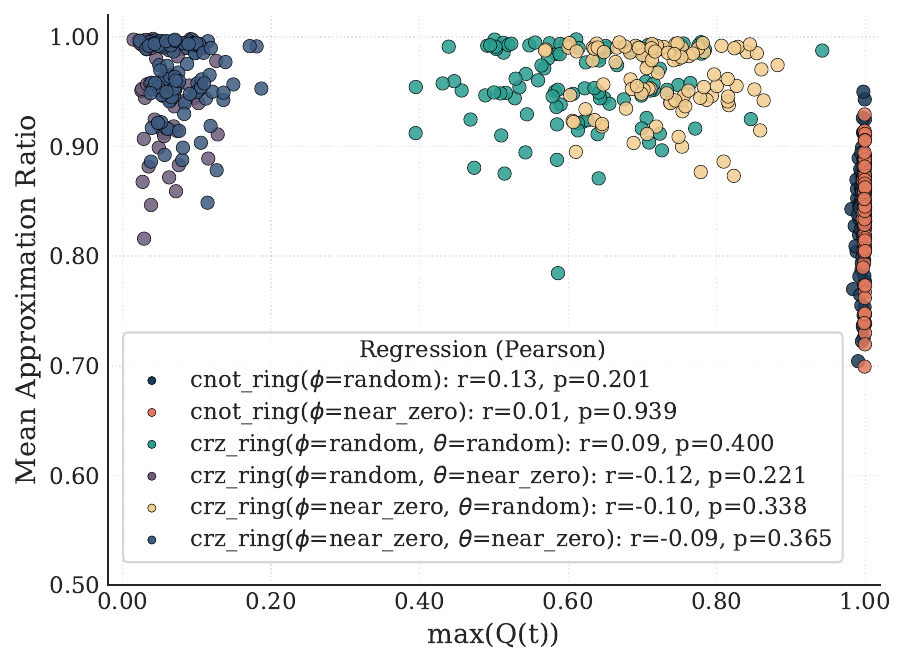}}
    \hfill
    \subfloat[Time-integrated $\mathrm{AUC}(Q)$ vs.\ quality.\label{fig:corr_auc}]{%
        \includegraphics[width=0.32\textwidth]{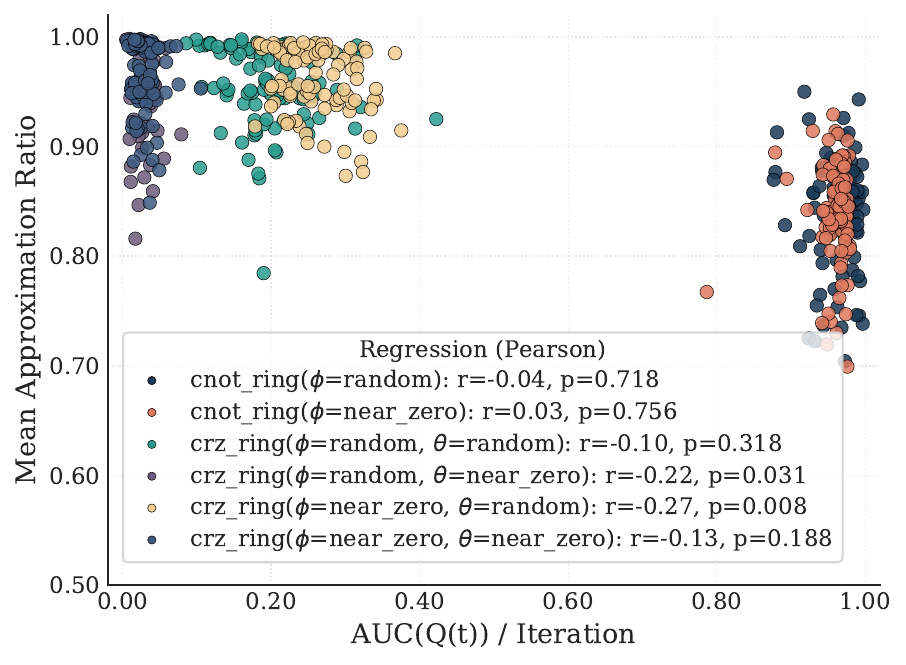}}
    \caption{Aggregate and instance-level solution quality. \textbf{(a)}~Distribution of approximation ratios per configuration. \textbf{(b)}~Comparison of mean approximation ratio and most probable bitstring quality for QAOA and the product-state ansatz; $\Delta$ denotes the gap. \textbf{(c)--(e)}~Instance-level correlation between entanglement summary statistics (final $Q$, peak $Q$, AUC of $Q(t)$) and approximation ratio for all HEA configurations; Pearson $r$ and $p$-values reported per configuration.}
    \label{fig:quality_combined}
\end{figure*}

Figure~\ref{fig:mean_approx_ratio} summarizes the approximation ratios across all optimization runs. A clear pattern emerges: all CRot-based variants and the product-state ansatz achieve mean approximation ratios between $0.95$ and $0.97$, well above the Goemans--Williamson bound. That all four CRot configurations converge to comparably high performance, regardless of initialization, is consistent with the trajectory analysis (cf.\ Figure~\ref{fig:q_trajectories_hea}), which showed that the optimizer drives entanglement toward zero in every case. The \textit{cnot\_ring} circuits, whose rigid entangling layer prevents this suppression, perform worst ($\alpha \approx 0.84$).

QAOA achieves mean approximation ratios of $\alpha \approx 0.85$--$0.88$, placing it below the CRot variants but near the \textit{cnot\_ring} circuits. This ranking must be interpreted with care: the approximation ratio averages over the full measurement distribution, whereas QAOA is designed to concentrate sampling probability on optimal or near-optimal bitstrings~\cite{farhi2014quantum}. Furthermore, QAOA operates with only $2p$ parameters ($6$--$14$ here) versus $\mathcal{O}(nL)$ for the HEAs.

Figure~\ref{fig:qaoa_bitstring} further substantiates this point by comparing the mean approximation ratio with the quality of the most probable bitstring. For QAOA, the gap is substantial ($\Delta = +0.104$ to $+0.136$): the most probable bitstring achieves $\alpha \approx 0.96$--$0.98$ across all depths, far exceeding the mean. The product-state ansatz shows no such discrepancy ($\Delta = +0.001$), indicating that the mean approximation ratio already captures its operational performance. Notably, QAOA's most probable bitstring quality surpasses that of the product-state ansatz at all depths, reversing the ranking implied by the mean approximation ratio. This confirms that QAOA's problem-structured entanglement builds constructive interference over multiple near-optimal solutions, a mechanism unavailable to separable ans\"atze, and that expectation-based evaluation systematically understates its operational performance.

\subsubsection{Instance-Level Entanglement--Performance Correlation}

Figure~\ref{fig:quality_combined}\textbf{(c)--(e)} moves from configuration-level averages to individual runs, correlating three entanglement summary statistics with approximation ratio. The clearest signal appears in the final-$Q$ panel~\textbf{(c)}: the CRot variants with randomly initialized entanglement parameters show significant negative correlations, confirming that instances on which the optimizer suppresses entanglement more yield better solutions. The near-zero-initialized CRot variants show weaker, non-significant correlations in this panel---as expected, since their final $Q$ values cluster near zero with insufficient variance to resolve a trend. Peak entanglement~\textbf{(d)} shows no significant correlation for any configuration, suggesting that transient entanglement excursions do not correlate with solution quality. The AUC measure~\textbf{(e)} recovers moderate negative correlations for the CRot variants. Overall, the instance-level data support the aggregate finding: lower entanglement is associated with better performance, but the effect is most detectable where the optimizer has sufficient dynamic range in $Q$ to reveal it.

\subsubsection{Entanglement Restriction through $\delta$ and $\rho$}
\label{sec:direct_sweep}
\begin{figure*}[t]
    \centering
    \subfloat[CNOT deletion factor~$\delta$.\label{fig:direct_sweep_deletion}]{%
        \includegraphics[width=0.33\textwidth]{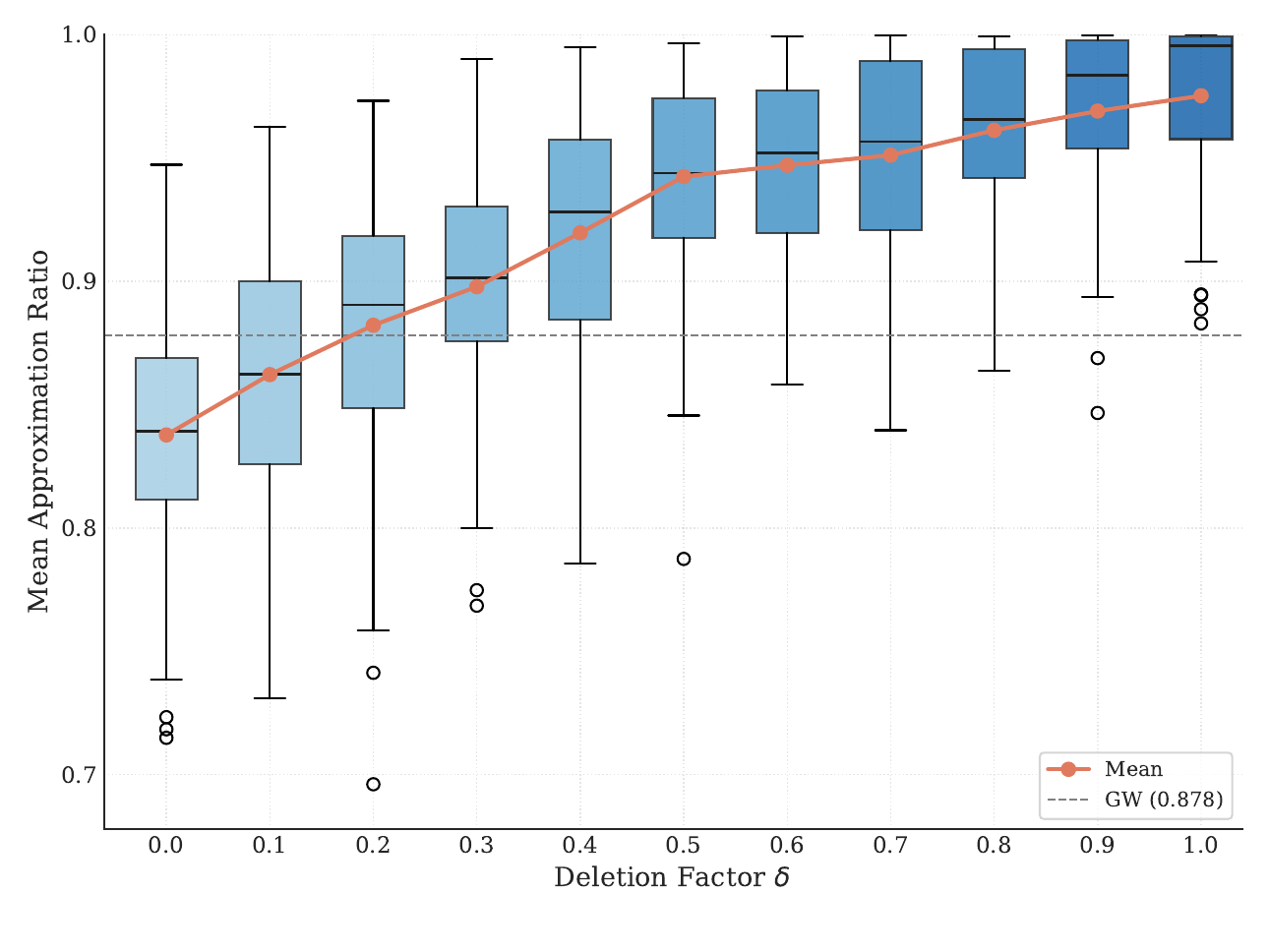}}
    \hfill
    \subfloat[CRot restriction factor~$\rho$.\label{fig:direct_sweep_restriction}]{%
        \includegraphics[width=0.33\textwidth]{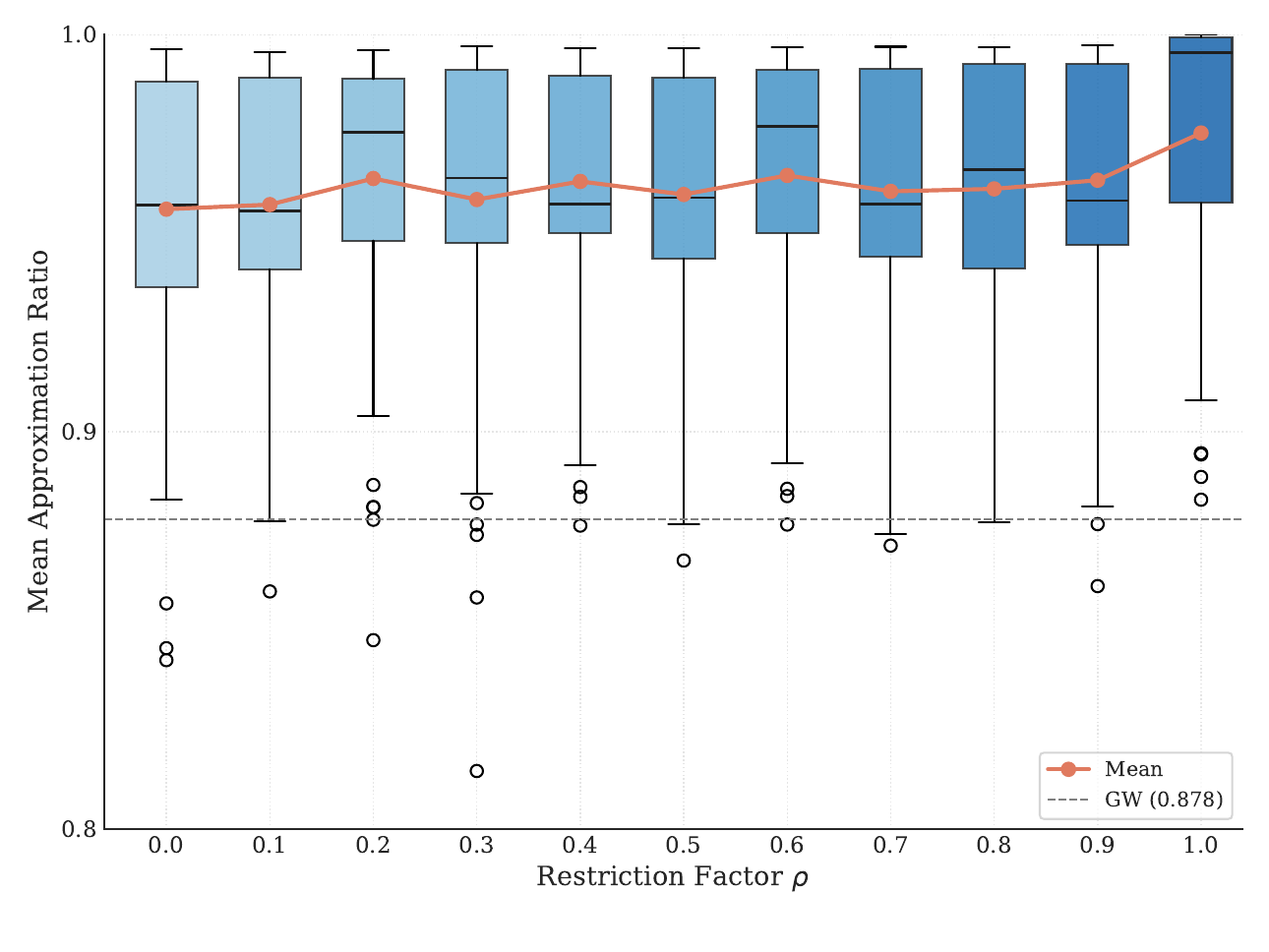}}
    \hfill
    \subfloat[Number of global optima.\label{fig:global_optima}]{%
        \includegraphics[width=0.33\textwidth]{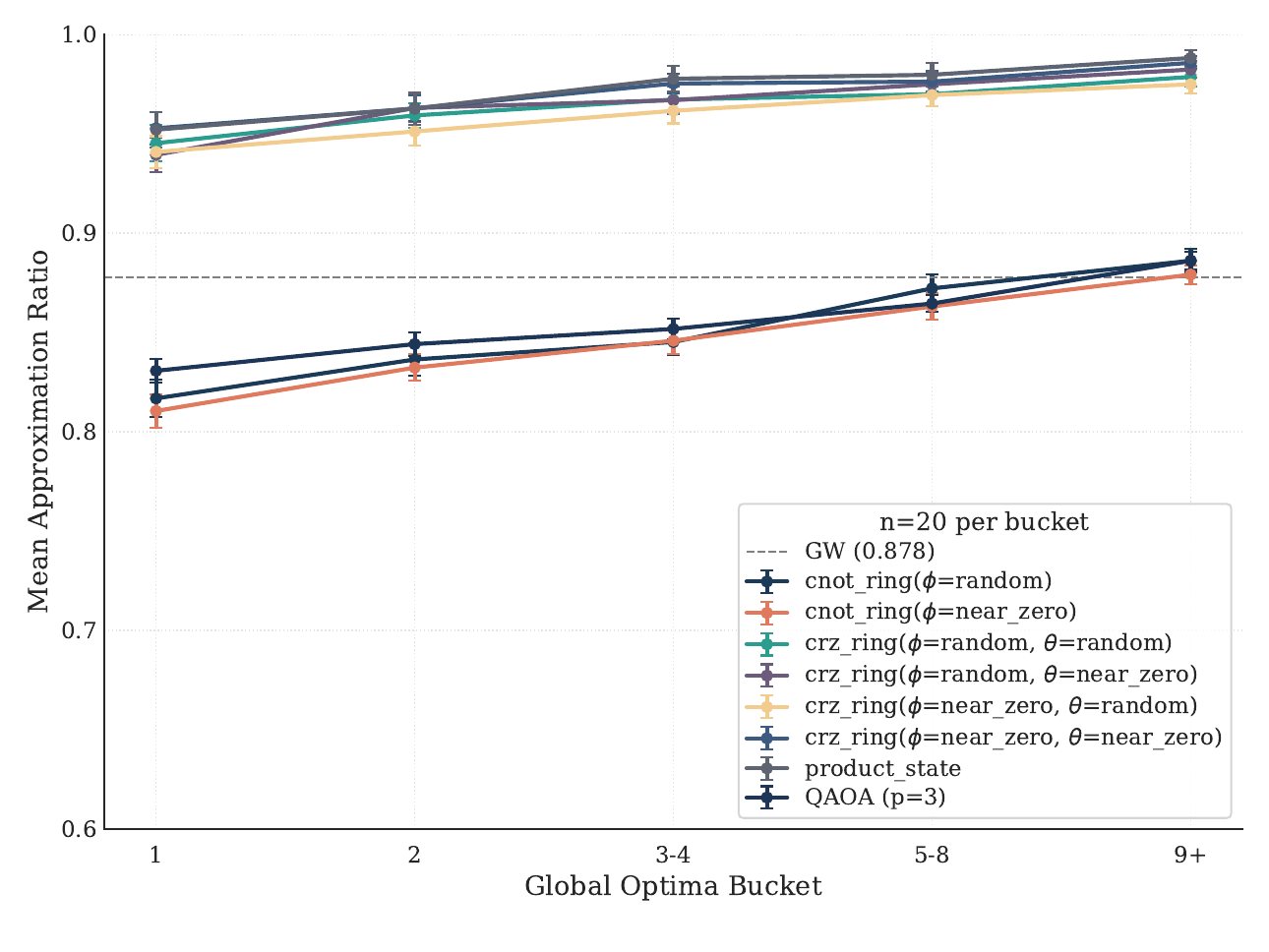}}
    \caption{Solution quality (approximation ratio) under controlled entanglement reduction and across problem structure. The dashed line marks the GW bound ($\alpha_{\mathrm{GW}} = 0.878$). \textbf{(a)}~Progressive CNOT deletion: performance increases monotonically with the fraction of removed gates. \textbf{(b)}~CRot parameter restriction: performance is approximately flat while entanglement remains saturated ($\rho \lesssim 0.7$) and rises once the angular budget restricts entanglement generation; at $\rho = 1$, deactivated CRZ gates are removed (see text). \textbf{(c)}~Approximation ratio by number of global optima; bucket size $=20$.}
    \label{fig:direct_sweep}
\end{figure*}

Having established in RQ1 that the deletion factor~$\delta$ and restriction factor~$\rho$ provide controlled modulation of circuit entanglement, we now investigate their direct impact on optimization performance. Our preceding results consistently indicate that lower entanglement is associated with higher solution quality. By actively restricting the entanglement budget to a low but nonzero level, we test whether there exists an optimal entanglement regime---a sweet spot---in which the circuit retains sufficient quantum correlations to, in principle, exceed classical simulability while avoiding the performance degradation caused by excessive problem-agnostic entanglement. We sweep both control parameters across their full range $[0,1]$ in increments of~$0.1$, evaluating each setting on all $20$ MaxCut instances with $5$ independent optimizer seeds ($100$ runs per setting). For the CNOT-deletion sweep, single-qubit parameters are initialized in the \textit{near\_zero} regime ($\boldsymbol{\phi}\sim\mathcal{U}[-0.1,+0.1]$). 
For the CRot-restriction sweep, single-qubit parameters are likewise initialized \textit{near\_zero}; the CRZ entangling parameters are initialized from $\mathcal{U}[-(1-\rho)\pi,\,(1-\rho)\pi]$ and constrained during optimization via COBYLA inequality box constraints. At $\rho=1$, we evaluate the corresponding product-state ansatz. Both sweeps use $500$ optimizer iterations per run. Results are shown in Fig.~\ref{fig:direct_sweep}\textbf{(a)--(b)}.

Under CNOT deletion (Fig.~\ref{fig:direct_sweep_deletion}), the mean approximation ratio increases monotonically from $\alpha \approx 0.84$ at $\delta = 0$ (full entangling circuit) to $\alpha \approx 0.97$ at $\delta = 0.9$. The product-state limit ($\delta = 1$) achieves even higher mean performance, confirming that no amount of problem-agnostic entanglement benefits the optimization. The relationship is remarkably smooth: each incremental removal of CNOT gates yields a measurable improvement, with no clear performance dip at any intermediate deletion level.

Under CRot restriction (Fig.~\ref{fig:direct_sweep_restriction}), the mean approximation ratio remains approximately constant at $\alpha \approx 0.95$ across $\rho \in [0.0,\, 0.7]$. Beyond $\rho \approx 0.7$, where the angular budget begins to curtail entanglement generation, the mean approximation ratio rises, culminating at $\rho = 1$ in the highest mean performance and the most concentrated distribution. At this product-state limit, the CRZ gates are removed from the circuit entirely (cf.\ Sec.~\ref{sec:entanglement_control}), eliminating dead-parameter flat directions and yielding a clean separable baseline. The correspondence between the entanglement saturation threshold and the onset of performance improvement provides direct causal evidence that the gains are driven by entanglement reduction rather than by incidental changes in parameter count or expressibility. Both sweep experiments confirm the central finding under controlled conditions: for MaxCut with HEAs, solution quality improves monotonically with entanglement suppression.

\subsubsection{Dependence on the Number of Global Optima}
Figure~\ref{fig:global_optima} illustrates the approximation ratio by the number of global optima of each MaxCut instance, probing whether ground-state degeneracy, i.e.\ how many bit strings achieve the maximum cut, affects the relative performance of entangled and separable ans\"atze. The approximation ratio generally slightly increases with the number of global optima across all configurations, which is expected: a larger set of degenerate ground states effectively enlarges the target basin in the solution landscape. We theorized that in principle entangled ans\"atze should benefit disproportionately from high degeneracy, as they alone can effectively represent superpositions over multiple optimal bit strings, occupying several solutions simultaneously. However, this theoretical advantage does not materialize in our data: separable circuits profit from the enlarged target basin similarly to their entangled counterparts, and the performance hierarchy established in the aggregate analysis remains largely preserved across all degeneracy buckets, with the low-entanglement CRot variants and the product-state baseline consistently leading. Overall, ground-state degeneracy does not appear to alter the relative benefit of entanglement in our experimental setting.

\subsubsection{Dependence on Graph Edge Probability}
We examined the approximation ratio as a function of the Erd\H{o}s--R\'enyi edge probability $p \in \{0.2, \ldots, 0.8\}$. The performance hierarchy established in the aggregate analysis is preserved across all density levels, with no significant interaction between edge probability and the relative ranking of configurations. All methods exhibit a mild upward trend with increasing density, somewhat more pronounced for the \textit{cnot\_ring} and QAOA variants, but the effect does not alter the conclusions drawn above. We therefore omit the figure for brevity.

\section{Discussion \& Limitations}
Our results converge on a central finding: for MaxCut solved via VQE and HEAs, problem-agnostic entanglement is not merely unhelpful, it is actively detrimental, and monotonically so.

\subsection{The Entanglement--Performance Relationship in HEAs}
Across all HEA configurations, lower entanglement is consistently associated with higher approximation ratios. The product-state ansatz achieves one of the best mean performance together with the CRot variants. The \textit{cnot\_ring} circuits perform worst among the HEA family. This monotonic relationship is robust: it holds across the full range of Erd\H{o}s--R\'enyi edge probabilities and is not modulated by ground-state degeneracy (Fig.~\ref{fig:global_optima}).

The entanglement trajectory analysis provides mechanistic insight into this relationship. When the ansatz grants the optimizer parametric control over the entangling gates (\textit{crz\_ring} circuits), the optimizer consistently suppresses entanglement during training. The \textit{crz\_ring($\boldsymbol{\phi}=\mathrm{random},\boldsymbol{\theta}=\mathrm{near\_zero}$)} configuration is particularly informative: its single-qubit rotations are already randomly initialized and span a diverse region of product-state space, yet the optimizer builds transient entanglement in the early iterations before suppressing it. Rather than a purposeful strategy, this transient rise in $Q$ is likely an artifact of COBYLA’s initial sampling phase, where the construction of the starting simplex necessitates exploratory steps into entangled regions of the Hilbert space. As the optimizer subsequently contracts its trust region and converges toward the classical ground state, it effectively sheds this incidental entanglement. The fact that a fully separable ansatz achieves the highest approximation ratio confirms that this involuntary excursion provides no functional benefit for solving diagonal Hamiltonians.

The underlying reason for this behavior is structural. The MaxCut Hamiltonian~\eqref{eq:maxcut_hamiltonian} is diagonal in the computational basis, and its ground state is a classical product state $|\mathbf{x}^*\rangle \in \{|0\rangle, |1\rangle\}^{\otimes n}$. Any entanglement present in the final variational state constitutes a deviation from the target. In \textit{cnot\_ring} circuits, this deviation is architecturally imposed: the entangling layer generates correlations along a fixed nearest-neighbor ring topology that bears no relationship to the problem graph $G$, and the optimizer is unable to eliminate them. In \textit{crz\_ring} circuits, the optimizer can suppress the entangling gates toward the identity, effectively recovering a product-state ansatz.

\subsection{QAOA: A Structurally Different Use of Entanglement}
The QAOA results present a striking contrast. Despite operating at a persistently high entanglement level ($Q \approx 0.85$), QAOA achieves approximation ratios that exceed those of the \textit{cnot\_ring} and, at higher depths, approach the Goemans--Williamson bound. Importantly, these figures understate QAOA's operational performance: the approximation ratio is based on the expectation value $\langle \psi | H_C | \psi \rangle$, which averages over the full measurement distribution and is therefore penalized by residual probability mass on suboptimal bitstrings. Since QAOA is designed to amplify the probability of optimal or near-optimal solutions, expectation-based approximation ratios underestimate the quality of the most likely sampled solution.

This observation appears to contradict the monotonic ``less entanglement is better'' relationship established for HEAs, but the contradiction is resolved by considering the \emph{structure} of the entanglement involved. The phase-separation unitary $e^{-i\gamma H_C}$ generates entanglement exclusively along the edges of the problem graph: every two-qubit interaction corresponds to a constraint in the MaxCut instance, so the correlations are structurally aligned with the cost function. In contrast, the \textit{cnot\_ring} is, in general, unrelated to the problem graph. The Meyer--Wallach measure $Q$ is blind to this distinction, it captures the \emph{amount} of entanglement but not its \emph{structure}, which is why two circuits at comparable $Q$ levels can yield vastly different optimization performance. Our results therefore suggest that the relevant divide is not between high and low entanglement, but between problem-aligned and problem-agnostic entanglement.

\subsection{Implications for Quantum Advantage}
The results raise a fundamental concern about the suitability of HEAs for combinatorial optimization. The argument proceeds in three steps.

First, quantum advantage over classical computation requires, at minimum, that the quantum circuit generates entanglement during its execution~\cite{gottesman1998heisenberg, jozsa2003role}. A circuit producing only product states at every time step is efficiently simulable classically: the optimization decomposes into $n$ independent single-qubit problems that can be solved on a classical computer in polynomial time.

Second, our results show that for MaxCut, problem-agnostic entanglement monotonically degrades HEA performance. The optimizer, when given the freedom to do so, drives the circuit toward a product-state solution. The best-performing HEA configuration is the product-state ansatz itself.

Third, combining these two observations leads to a dilemma: the operating regime where HEAs perform best for MaxCut (low or zero entanglement) is precisely the regime that is efficiently simulable classically. Adding problem-agnostic entanglement does not improve performance---it degrades it. This empirical dilemma is supported by recent rigorous theoretical bounds: Leone et al.~\cite{leone2024practical} show that shallow HEAs, when applied to separable or more generally low-bond-dimension input states (as is typical in VQA and combinatorial optimization settings), are efficiently simulable with polynomial overhead via tensor-network methods. Thus, in this shallow-HEA regime, no quantum advantage is expected. Increasing depth to move beyond this regime does not automatically resolve the issue, as trainability can then deteriorate through concentration effects and barren-plateau phenomena~\cite{mcclean2018barren, ortiz2021entanglement, leone2024practical}.

This does not imply that variational quantum optimization is inherently incapable of delivering quantum advantage for combinatorial problems. The QAOA comparison demonstrates that problem-structured entanglement can coexist with good optimization performance, and theoretical results guarantee that QAOA converges to the optimal solution as $p \to \infty$~\cite{farhi2014quantum}.

\subsection{Limitations}
Several limitations of this study should be acknowledged.

\textit{Problem size.} All experiments are conducted at $n = 12$ qubits, where the Hilbert space ($\mathrm{dim} = 2^{12} = 4096$) is small enough that a product-state ansatz with $\mathcal{O}(n)$ parameters can effectively explore the solution landscape. At larger scales, the combinatorial explosion of the solution space may alter the entanglement--performance relationship: product-state optimization might encounter coordination and exploration difficulties that some degree of entanglement could alleviate. The qualitative direction of our findings, that the optimizer suppresses problem-agnostic entanglement when possible, is expected to persist, but the quantitative performance gaps may shift. Notably, the scaling outlook is asymmetric: Ortiz-Marrero et al.~\cite{ortiz2021entanglement} showed that ansätze whose entanglement satisfies volume-law scaling can suffer from entanglement-induced barren plateaus, exponentially vanishing gradients that arise as a direct consequence of excessive entanglement, independent of circuit depth. Since our high-entanglement HEA configurations already operate at high $Q$-values, this trainability bottleneck would compound the performance penalty we observe, likely widening rather than narrowing the gap between high- and low-entanglement configurations at scale.

\textit{Problem class.} Our analysis is restricted to MaxCut, a problem with a diagonal Hamiltonian and product-state ground states. Problems whose ground states are genuinely entangled, such as those arising in quantum chemistry or condensed matter physics, constitute a fundamentally different setting where entanglement in the ansatz is not merely a transient resource but a necessary feature of the target state. Our conclusions about the detrimental role of problem-agnostic entanglement should not be extrapolated to these domains as those problems are fundamentally different.

\textit{Classical optimizer.} Our results are obtained with a single classical optimizer. Different optimizers may navigate the entanglement landscape differently, potentially altering the observed trajectories and convergence behavior. 

\subsection{From Correlation to Controlled Evidence}
The direct sweep experiments (Fig.~\ref{fig:direct_sweep}) elevate the entanglement--performance relationship from a correlational observation to a controlled experimental finding. By holding all circuit parameters except the entanglement control knob constant, we isolate entanglement as the causal variable. The monotonic improvement under CNOT deletion and the plateau--onset structure under CRot restriction mostly rule out confounding explanations based on parameter count or expressibility changes. More fundamentally, no amount of careful dosing can compensate for the structural mismatch: HEA entangling topologies are dictated by hardware connectivity rather than by~$H_C$, and restricting their strength merely mitigates a design flaw rather than correcting it.

\section{Conclusion}
We investigated the role of entanglement in variational quantum optimization for MaxCut by introducing two mechanisms that provide controlled, monotonic modulation of HEAs entanglement. Our central finding is that problem-agnostic entanglement, as generated by standard HEAs, is monotonically detrimental: the optimizer suppresses it when given control, and a fully separable ansatz outperforms all entangled HEA configurations. In contrast, QAOA maintains high entanglement yet achieves competitive solution quality, while its correlations are structurally derived from the problem Hamiltonian. This indicates that entanglement structure, not quantity, determines its utility in variational optimization.

These results challenge a widespread and largely unexamined practice in the variational quantum computing community: the default inclusion of problem-agnostic entangling layers. For the broad class of combinatorial problems governed by diagonal Hamiltonians, our findings indicate that HEAs are fundamentally mismatched with the target state, and that their best operating regime is classically simulatable. This work provides empirical grounding for a shift toward problem-structured ansatz design and underscores the need to move beyond hardware convenience as the guiding principle for variational circuit construction.

\section*{Acknowledgement}
This paper was partially funded by the German Federal Ministry of Education and Research through the funding program “quantum technologies -- from basic research to market” (contract number: 13N16196). Generative AI was utilized to generate sections of this Work, including text, tables, graphs, code, citations, etc.

\bibliographystyle{IEEEtran}
\bibliography{references}

@article{gross2009most,
  title={Most quantum states are too entangled to be useful as computational resources},
  author={Gross, David and Flammia, Steve T and Eisert, Jens},
  journal={Physical review letters},
  volume={102},
  number={19},
  pages={190501},
  year={2009},
  publisher={APS}
}

@article{diez2021quantum,
  title={Quantum variational optimization: The role of entanglement and problem hardness},
  author={D{\'\i}ez-Valle, Pablo and Porras, Diego and Garc{\'\i}a-Ripoll, Juan Jos{\'e}},
  journal={Physical Review A},
  volume={104},
  number={6},
  pages={062426},
  year={2021},
  publisher={APS}
}

@article{kim2022quantum,
  title={Quantum energy landscape and circuit optimization},
  author={Kim, Joonho and Oz, Yaron},
  journal={Physical Review A},
  volume={106},
  number={5},
  pages={052424},
  year={2022},
  publisher={APS}
}

@article{nakhl2024calibrating,
  title={Calibrating the role of entanglement in variational quantum circuits},
  author={Nakhl, Azar C and Quella, Thomas and Usman, Muhammad},
  journal={Physical Review A},
  volume={109},
  number={3},
  pages={032413},
  year={2024},
  publisher={APS}
}

@article{wang2024entanglement,
  title={Entanglement-variational hardware-efficient ansatz for eigensolvers},
  author={Wang, Xin and Qi, Bo and Wang, Yabo and Dong, Daoyi},
  journal={Physical Review Applied},
  volume={21},
  number={3},
  pages={034059},
  year={2024},
  publisher={APS}
}

@article{meyer2002global,
  title={Global entanglement in multiparticle systems},
  author={Meyer, David A and Wallach, Nolan R},
  journal={Journal of Mathematical Physics},
  volume={43},
  number={9},
  pages={4273--4278},
  year={2002},
  publisher={American Institute of Physics}
}

@article{brennen2003observable,
  title={An observable measure of entanglement for pure states of multi-qubit systems},
  author={Brennen, Gavin K},
  journal={Quantum Information \& Computation},
  volume={3},
  number={6},
  pages={619--626},
  year={2003},
  publisher={Rinton Press, Incorporated Paramus, NJ}
}

@article{goemans1995improved,
  title={Improved approximation algorithms for maximum cut and satisfiability problems using semidefinite programming},
  author={Goemans, Michel X and Williamson, David P},
  journal={Journal of the ACM (JACM)},
  volume={42},
  number={6},
  pages={1115--1145},
  year={1995},
  publisher={ACM New York, NY, USA}
}

@article{peruzzo2014variational,
  title={A variational eigenvalue solver on a photonic quantum processor},
  author={Peruzzo, Alberto and McClean, Jarrod and Shadbolt, Peter and Yung, Man-Hong and Zhou, Xiao-Qi and Love, Peter J and Aspuru-Guzik, Al{\'a}n and O’brien, Jeremy L},
  journal={Nature communications},
  volume={5},
  number={1},
  pages={4213},
  year={2014},
  publisher={Nature Publishing Group UK London}
}

@article{farhi2014quantum,
  title={A quantum approximate optimization algorithm},
  author={Farhi, Edward and Goldstone, Jeffrey and Gutmann, Sam},
  journal={arXiv preprint arXiv:1411.4028},
  year={2014}
}

@article{kandala2017hardware,
  title={Hardware-efficient variational quantum eigensolver for small molecules and quantum magnets},
  author={Kandala, Abhinav and Mezzacapo, Antonio and Temme, Kristan and Takita, Maika and Brink, Markus and Chow, Jerry M and Gambetta, Jay M},
  journal={nature},
  volume={549},
  number={7671},
  pages={242--246},
  year={2017},
  publisher={Nature Publishing Group}
}

@inproceedings{rohe2024questionable,
  title={The questionable influence of entanglement in quantum optimisation algorithms},
  author={Rohe, Tobias and Schuman, Dani{\"e}lle and N{\"u}blein, Jonas and S{\"u}nkel, Leo and Stein, Jonas and Linnhoff-Popien, Claudia},
  booktitle={2024 IEEE International Conference on Quantum Computing and Engineering (QCE)},
  volume={1},
  pages={1497--1503},
  year={2024},
  organization={IEEE}
}

@article{amaro2022filtering,
  title={Filtering variational quantum algorithms for combinatorial optimization},
  author={Amaro, David and Modica, Carlo and Rosenkranz, Matthias and Fiorentini, Mattia and Benedetti, Marcello and Lubasch, Michael},
  journal={Quantum Science \& Technology},
  volume={7},
  number={1},
  pages={015021},
  year={2022},
  publisher={IOP Publishing}
}

@article{liu2022layer,
  title={Layer VQE: A variational approach for combinatorial optimization on noisy quantum computers},
  author={Liu, Xiaoyuan and Angone, Anthony and Shaydulin, Ruslan and Safro, Ilya and Alexeev, Yuri and Cincio, Lukasz},
  journal={IEEE Transactions on Quantum Engineering},
  volume={3},
  pages={1--20},
  year={2022},
  publisher={IEEE}
}

@article{miki2022variational,
  title={Variational ansatz preparation to avoid cnot-gates on noisy quantum devices for combinatorial optimizations},
  author={Miki, Tsukasa and Okita, Ryo and Shimada, Moe and Tsukayama, Daisuke and Shirakashi, Jun-ichi},
  journal={AIP Advances},
  volume={12},
  number={3},
  year={2022},
  publisher={AIP Publishing}
}

@article{barkoutsos2020improving,
  title={Improving variational quantum optimization using CVaR},
  author={Barkoutsos, Panagiotis Kl and Nannicini, Giacomo and Robert, Anton and Tavernelli, Ivano and Woerner, Stefan},
  journal={Quantum},
  volume={4},
  pages={256},
  year={2020},
  publisher={Verein zur F{\"o}rderung des Open Access Publizierens in den Quantenwissenschaften}
}

@article{nannicini2019performance,
  title={Performance of hybrid quantum-classical variational heuristics for combinatorial optimization},
  author={Nannicini, Giacomo},
  journal={Physical Review E},
  volume={99},
  number={1},
  pages={013304},
  year={2019},
  publisher={APS}
}

@article{jozsa2003role,
  title={On the role of entanglement in quantum-computational speed-up},
  author={Jozsa, Richard and Linden, Noah},
  journal={Proceedings of the Royal Society of London. Series A: Mathematical, Physical and Engineering Sciences},
  volume={459},
  number={2036},
  pages={2011--2032},
  year={2003},
  publisher={The Royal Society}
}

@article{gottesman1998heisenberg,
  title={The Heisenberg representation of quantum computers},
  author={Gottesman, Daniel},
  journal={arXiv preprint quant-ph/9807006},
  year={1998}
}

@inproceedings{turati2023benchmarking,
  title={Benchmarking adaptative variational quantum algorithms on qubo instances},
  author={Turati, Gloria and Dacrema, Maurizio Ferrari and Cremonesi, Paolo},
  booktitle={2023 IEEE International Conference on Quantum Computing and Engineering (QCE)},
  volume={1},
  pages={407--413},
  year={2023},
  organization={IEEE}
}

@article{dupont2022entanglement,
  title={Entanglement perspective on the quantum approximate optimization algorithm},
  author={Dupont, Maxime and Didier, Nicolas and Hodson, Mark J and Moore, Joel E and Reagor, Matthew J},
  journal={Physical Review A},
  volume={106},
  number={2},
  pages={022423},
  year={2022},
  publisher={APS}
}

@article{ortiz2021entanglement,
  title={Entanglement-induced barren plateaus},
  author={Ortiz Marrero, Carlos and Kieferov{\'a}, M{\'a}ria and Wiebe, Nathan},
  journal={PRX quantum},
  volume={2},
  number={4},
  pages={040316},
  year={2021},
  publisher={APS}
}

@article{leone2024practical,
  title={On the practical usefulness of the hardware efficient ansatz},
  author={Leone, Lorenzo and Oliviero, Salvatore FE and Cincio, Lukasz and Cerezo, Marco},
  journal={Quantum},
  volume={8},
  pages={1395},
  year={2024},
  publisher={Verein zur F{\"o}rderung des Open Access Publizierens in den Quantenwissenschaften}
}

@incollection{Karp1972Reducibility,
  author    = {Richard M. Karp},
  title     = {Reducibility among combinatorial problems},
  booktitle = {Complexity of Computer Computations},
  editor    = {Raymond E. Miller and James W. Thatcher and Jean D. Bohlinger},
  publisher = {Springer},
  address   = {Boston, MA},
  pages     = {85--103},
  year      = {1972},
  series    = {The IBM Research Symposia Series},
  doi       = {10.1007/978-1-4684-2001-2_9}
}

@article{mcclean2018barren,
  title={Barren plateaus in quantum neural network training landscapes},
  author={McClean, Jarrod R and Boixo, Sergio and Smelyanskiy, Vadim N and Babbush, Ryan and Neven, Hartmut},
  journal={Nature communications},
  volume={9},
  number={1},
  pages={4812},
  year={2018},
  publisher={Nature Publishing Group UK London}
}

\vspace{12pt}

\end{document}